\documentclass[twocolumn]{jpsj2} 
\title{Quantum Turbulence}

\author{Makoto \textsc{Tsubota}$^{1}$ \thanks{E-mail: tsubota@sci.osaka-cu.ac.jp} \thanks{Present address:  Department of Physics, Osaka City University, Osaka 558-8585, Japan.}}

\inst{$^{1}$ Department of Physics, Osaka City University, Japan}

\abst{The present article reviews the recent developments in the physics of quantum turbulence. Quantum turbulence (QT) was discovered in superfluid $^4$He in the 1950s, and the research has tended toward a new direction
since the mid 90s.  The similarities and differences between quantum and classical turbulence have become an important area of research.  QT is comprised of quantized vortices that are definite topological defects, being expected to yield a model of turbulence that is much simpler than the classical model. The general introduction of the issue and a brief review on classical turbulence are followed by a description of the dynamics of quantized vortices.  Then, we discuss the energy spectrum of QT at very low temperatures. At low wavenumbers, the energy is transferred through the Richardson cascade of quantized vortices, and the spectrum obeys the Kolmogorov law, which is the most important statistical law in turbulence; this classical region shows the similarity to conventional turbulence. At higher wavenumbers, the energy is transferred by the Kelvin-wave cascade on each vortex. This quantum regime depends strongly on the nature of each quantized vortex. The possible dissipation mechanism is discussed. Finally, important new experimental studies, which include investigations into temperature-dependent transition to QT, dissipation at very low temperatures, QT created by vibrating structures, and visualization of QT, are reviewed.  The present article concludes with a brief look at QT in atomic Bose-Einstein condensates.}

\kword{Quantum turbulence, Quantized vortex, Superfluid, Bose-Einstein condensation}

\begin{document}
\maketitle

\section{Introduction} 
For high-velocity or high Reynolds number flows, the flow is generally turbulent. Turbulence has been investigated not only in basic science research, such as physics and mathematics research, but also in applied sciences, such as fluid engineering and aeronautics. Although turbulence has been studied intensely in a number of fields, it is still not yet well understood. This is chiefly because turbulence is a complicated dynamical phenomenon with strong nonlinearity that is quite different from an equilibrium state. Vortices may be key in understanding turbulence. For example, Leonardo da Vinci observed the turbulent flow of water and drew many sketches showing that turbulence had a structure comprised of vortices of different sizes. However, vortices are not well-defined for a typical classical fluid, and the relationship between turbulence and vortices remains unclear. 

Independently of these studies in classical fluid dynamics,  
turbulence in superfluid helium
has been studied in the field of low-temperature physics.
Liquid $^4$He enters a superfluid state below the $\lambda$ point (2.17 K) with
Bose--Einstein condensation of the $^4$He atoms \cite{TilleyTilley}. 
Characteristic phenomena of superfluidity were discovered experimentally
in the 1930s by Kapitza et al.
The hydrodynamics of superfluid helium is well described by the two-fluid model,
for which the system consists of an inviscid superfluid (density $\rho_s$) and 
a viscous normal fluid (density $\rho_n$) with two independent velocity fields $\mathbf{v}_s$ and $\mathbf{v}_n$. 
The mixing ratio of the two fluids depends on temperature. 
As the temperature is reduced below the $\lambda$ point, the ratio of 
the superfluid component increases, and the entire fluid becomes
a superfluid below approximately 1 K.
The Bose-condensed system exhibits the macroscopic wave function
$\Psi(\mathbf{x},t)=|\Psi(\mathbf{x},t)| e^{i \theta(\mathbf{x},t)}$ as an order parameter. 
The superfluid velocity field is given by $\mathbf{v}_s=(\hbar/m) \nabla \theta$ with boson mass $m$,
representing the potential flow. 
Since the macroscopic wave function should be single-valued for the space coordinate $\mathbf{x}$, the circulation $\Gamma = \oint \mathbf{v} \cdot d\mathbf{\ell}$ for an arbitrary closed loop in the fluid is quantized by the quantum $\kappa=h/m$.  A vortex with such quantized circulation is called a quantized vortex. Any rotational motion of a superfluid is sustained only
by quantized vortices.

A quantized vortex is a topological defect characteristic of a Bose--Einstein condensate and is different from a vortex in a classical viscous fluid. First, the circulation is quantized, which is contrary to a classical vortex that can have any value of circulation. Second, a quantized vortex is a vortex of inviscid superflow. Thus, it cannot decay by the viscous diffusion of vorticity that occurs in a classical fluid.  Third, the core of a quantized vortex is very thin, on the order of the coherence length, which is only a few angstroms in superfluid $^4$He. Since the vortex core is very thin and does not decay by diffusion, it is always possible to identify the position of a quantized vortex in the fluid. These properties make a quantized vortex more stable and definite than a classical vortex. 

Early experimental studies on superfluid turbulence focused primarily on thermal counterflow, in which the normal fluid and superfluid flow in opposite directions. The flow is driven by an injected heat current, and it was found that the superflow becomes dissipative when the relative velocity between the two fluids exceeds a critical value \cite{GorterMellink}. Feynman proposed that this is a superfluid turbulent state consisting of a tangle of quantized vortices \cite{Feynman}. Vinen later confirmed Feynman's findings experimentally by showing that the dissipation comes from the mutual friction between vortices and the normal flow \cite{Vinen57a, Vinen57b, Vinen57c}. Subsequently, several experimental studies have examined superfluid turbulence (ST) in thermal counterflow systems and have revealed a variety of physical phenomenon\cite{Tough82}. Since the dynamics of quantized vortices is nonlinear and non-local, it has not been easy to understand vortex dynamics observations quantitatively. Schwarz clarified the picture of ST consisting of tangled vortices by a numerical simulation of the quantized vortex filament model in the thermal counterflow \cite{Schwarz85, Schwarz88}. However, since the thermal counterflow has no analogy to conventional fluid dynamics, this study was not helpful in clarifying the relationship between ST and classical turbulence (CT). Superfluid turbulence is often called quantum turbulence (QT), which emphasizes the fact that it is comprised of quantized vortices.

Comparing QT and CT reveals definite differences, which demonstrates the importance of studying QT. Turbulence in a classical viscous fluid appears to be comprised of vortices, as pointed out by Da Vinci. However, these vortices are unstable, and appear and disappear repeatedly. Moreover, the circulation is not conserved and is not identical for each vortex. Quantum turbulence consists of a tangle of quantized vortices that have the same conserved circulation. Thus, QT is an easier system to study than CT and has a much simpler model of turbulence than CT. 

Based on this consideration, QT research has tended toward a new direction
since the mid 90s. A chief interest is to understand the relationship between
QT and CT \cite{Vinen02, Vinen06}.  
The energy spectrum of fully developed classical turbulence is known to obey the Kolmogorov law in an inertial range. The energy transfer in an inertial range is believed to be sustained by the Richardson cascade process, in which large eddies are broken up self-similarly to smaller eddies. Recent experimental and numerical studies support the Kolmogorov spectrum in QT. Another important problem is the dissipative mechanism in QT.  At a finite temperature the mutual friction works as the dissipative mechanism. However, it is not so easy to understand what kind of mechanism causes dissipation at very low temperatures in which the normal fluid component is negligible.

This article reviews recent research on QT. After reviewing classical turbulence in Section 2, Section 3 describes the dynamics of quantized vortices, which is key in understanding QT. Section 4 summarizes the motivation and the problems of recent research on quantum turbulence. In Section 5, we discuss the issue of the energy spectra and a possible dissipation mechanism at very low temperatures. Section 6 describes recent important experimental studies, including research into the temperature-dependent transition to QT, dissipation at very low temperatures, QT created by vibrating structures, and visualization of QT. Section 7 presents a summary and briefly addresses QT in atomic Bose-Einstein condensates.  

\section{Classical turbulence}
Before considering QT, we briefly review classical fluid dynamics and the statistical properties of CT \cite{Frisch}.
Classical viscous fluid dynamics is described by the Navier--Stokes equation:
\begin{equation}
\frac{\partial}{\partial t} \mathbf{v}(\mathbf{x},t) + \mathbf{v}(\mathbf{x},t) \cdot \nabla \mathbf{v}(\mathbf{x},t) = - \frac{1}{\rho} \nabla P(\mathbf{x},t) + \nu \nabla^2 \mathbf{v}(\mathbf{x},t),
\label{eq-Navier-Stokes}
\end{equation}
where $\mathbf{v}(\mathbf{x},t)$ is the velocity of the fluid, $P(\mathbf{x},t)$ is the pressure, $\rho$ is the density of the fluid, and $\nu$ is the kinematic viscosity. The flow of this fluid can be characterized by the ratio of the second term of the left-hand side of Eq. (\ref{eq-Navier-Stokes}), hereinafter referred to as the inertial term, to the second term of the right-hand side, hereinafter called the viscous term. This ratio is the Reynolds number $R=\bar{v}D/\nu$, where $\bar{v}$ and $D$ are the characteristic velocity of the flow and the characteristic scale, respectively. When $\bar{v}$ increases to allow the Reynolds number to exceed a critical value, the system changes from a laminar state  to a turbulent state, in which the flow is highly complicated and contains many eddies.

Such turbulent flow is known to show characteristic statistical behavior \cite{Kolmogorov41a, Kolmogorov41b}. 
We assume a steady state of fully developed turbulence of an incompressible classical fluid. 
The energy is injected into the fluid at a rate of $\varepsilon$, the scale of which is comparable to the system size $D$ in the energy-containing range. In the inertial range, this energy is transferred to smaller scales without being dissipated. In this range, the system is locally homogeneous and isotropic, which leads to the statistics of the energy spectrum known as the Kolmogorov law:
\begin{equation} E(k)=C\varepsilon^{2/3}k^{-5/3}.\label{eq-Kolmogorov} \end{equation}
Here, the energy spectrum $E(k)$ is defined as $E=\int d \mathbf{k} E(k)$, where $E$ is the kinetic energy per unit mass and $k$ is the wavenumber from the Fourier transformation of the velocity field. The spectrum of Eq. (\ref{eq-Kolmogorov}) is easily derived by assuming that $E(k)$ is locally determined by only the energy flux $\varepsilon$ and $k$. The energy transferred to smaller scales in the energy-dissipative range is dissipated at the Kolmogorov wavenumber $k_K=(\epsilon/\nu^3)^{1/4}$ through the viscosity of the fluid with dissipation rate $\varepsilon$ in Eq. (\ref{eq-Kolmogorov}), which is equal to the energy flux $\Pi$ in the inertial range. The Kolmogorov constant $C$ is a dimensionless parameter of order unity. The Kolmogorov spectrum is confirmed experimentally and numerically in turbulence at high Reynolds numbers. 
The inertial range is thought to be sustained by the self-similar Richardson cascade in which large eddies are broken up into smaller eddies through many vortex reconnections. In CT, however, the Richardson cascade is not completely understood because it is impossible to definitely identify each eddy. 
The Kolmogorov spectrum is based on the assumption that the turbulence is homogeneous and isotropic.
However, actual turbulence is not necessarily homogeneous or isotropic, and so 
the energy spectrum deviates from the Kolmogorov form.
This phenomenon is called intermittency, and is an important problem 
in modern fluid dynamics \cite{Frisch}.
Intermittency is closely related to the coherent structure, which may be represented by
vortices. Research into QT could also shed light on this issue.

\section{Dynamics of quantized vortices}
As described in Section 1, most experimental studies on superfluid turbulence have examined thermal counterflow. However, the nonlinear and non-local dynamics of vortices have delayed progress in the microscopic understanding of the vortex tangle. Schwarz overcame these difficulties \cite{Schwarz85, Schwarz88} by developing a direct numerical simulation of vortex dynamics connected with dynamical scaling analysis, enabling the calculation of physical quantities such as the vortex line density, anisotropic parameters, and mutual friction force. The observable quantities obtained by Schwarz agree well with some typical experimental results.  

Two formulations are generally available for studying the dynamics of quantized vortices.  One is the vortex filament model and the other is the Gross--Pitaevskii (GP) model. We will briefly review these two formulations.

\subsection{Vortex filament model}
As described in Section 1, a quantized vortex has quantized circulation. The vortex core is extremely thin, usually much smaller than other characteristic scales in vortex motion. These properties allow a quantized vortex to be represented as a vortex filament. In classical fluid dynamics \cite{Saffman}, the vortex filament model is a convenient idealization. However, the vortex filament model is accurate and realistic for a quantized vortex in superfluid helium.

The vortex filament formulation represents a quantized vortex as a filament passing through the fluid, having a definite direction corresponding to its vorticity. Except for the thin core region, the superflow velocity field has a classically well-defined meaning and can be described by ideal fluid dynamics. The velocity at a point $\mathbf{r}$ due to a filament is given by the Biot--Savart expression:
\begin{equation}
\mathbf{v}_{s} (\mathbf{r} )=\frac{\kappa}{4\pi}\int_{\cal L} \frac{(\mathbf{s}_1 - \mathbf{r}) \times
d\mathbf{s}_1}{|\mathbf{s}_1-\mathbf{r}|^3},
\label{BS}
\end{equation}
where $\kappa$ is the quantum of circulation. The filament is represented by the parametric form $\mathbf{s} = \mathbf{s}(\xi, t)$ with the one-dimensional coordinate $\xi$ along the filament. The vector $\mathbf{s}_1$ refers to a point on the filament, and the integration is taken along the filament. Helmholtz's theorem for a perfect fluid states that the vortex moves with the superfluid velocity. Calculating the velocity $\mathbf{v}_{s}$ at a point $\mathbf{r}=\mathbf{s}$ on the filament causes the integral to diverge as $\mathbf{s}_1 \rightarrow \mathbf{s}$. To avoid this divergence, we separate the velocity $\dot{\mathbf{s}}$ of the filament at the point $\mathbf{s}$ into two components \cite{Schwarz85}:
\begin{equation}
\dot{\mathbf{s}} =\frac{\kappa}{4\pi}\mathbf{s}' \times \mathbf{s}'' \ln \left(
\frac{2(\ell_+ \ell_-)^{1/2}}
{e^{1/4} a_0}\right) + \frac{\kappa}{4\pi}\int_{\cal L}^{'} \frac{(\mathbf{s}_1 -
\mathbf{r})
\times d\mathbf{s}_1}{|\mathbf{s}_1-\mathbf{r}|^3}.
\label{sdot}
\end{equation}
The first term is the localized induction field arising from a curved line element acting on itself, and $\ell_+$ and $\ell_-$ are the lengths of the two adjacent line elements after discretization, separated by the point $\mathbf{s}$. The prime denotes differentiation with respect to the arc length $\xi$. The mutually perpendicular vectors $\mathbf{s}'$, $\mathbf{s}''$, and $\mathbf{s}' \times \mathbf{s}''$ point along the tangent, the principal normal, and the binormal, respectively, at the point $\mathbf{s}$, and their respective magnitudes are 1, $R^{-1}$, and $R^{-1}$, with the local radius $R$ of curvature. The parameter $a_0$ is the cutoff corresponding to the core radius. Thus, the first term represents the tendency to move the local point $\mathbf{s}$ along the binormal direction with a velocity inversely proportional to $R$. The second term represents the non-local field obtained by integrating the integral of Eq. (\ref{BS}) along the rest of the filament, except in the neighborhood of $\mathbf{s}$. 

Neglecting the non-local terms and replacing Eq. (\ref{sdot}) by $\dot{\mathbf{s}} = \beta \mathbf{s}' \times \mathbf{s}''$ is referred to as the localized induction approximation (LIA). Here, the coefficient $\beta$ is defined by $\beta = (\kappa/4\pi) \ln \left( c \langle R \rangle/a_0 \right)$, where $c$  is a constant of order 1 and $(\ell_+\ell_-)^{1/2}$ is replaced by the mean radius of curvature $\langle R \rangle$ along the length of the filament.  This approximation is believed to be effective for analyzing isotropic dense tangles due to cancellations between non-local contributions. 

A better understanding of vortices in a real system results is obtained when boundaries are included in the analysis. For this purpose, a boundary-induced velocity field $\mathbf{v}_{s,b}$ is added to $\mathbf{v}_{s}$, so that the superflow can satisfy the boundary condition of an inviscid flow, that is, that the normal component of the velocity should disappear at the boundaries. To allow for another, presently unspecified, applied field, we include $\mathbf{v}_{s,a}$. Hence, the total velocity $\dot{\mathbf{s}}_0$ of the vortex filament without dissipation is 
\begin{eqnarray}
\dot{\mathbf{s}}_0 =\frac{\kappa}{4\pi}\mathbf{s}' \times \mathbf{s}'' \ln \left( \frac{2(\ell_+
\ell_-)^{1/2}}{e^{1/4} a_0}\right) + \frac{\kappa}{4\pi}\int_{\cal L}^{'}
\frac{(\mathbf{s}_1 - \mathbf{r}) \times d\mathbf{s}_1}{|\mathbf{s}_1-\mathbf{r}|^3} \nonumber\\  
+\mathbf{v}_{s,b}(\mathbf{s})+ \mathbf{v}_{s,a}(\mathbf{s}).
\label{s0dot}
\end{eqnarray}
\noindent
At finite temperatures, it is necessary to take into account the mutual friction between the vortex core and the normal flow $\mathbf{v}_n$. 
Including this term, the velocity of $\mathbf{s}$ is given by 
\begin{equation}
\dot{\mathbf{s}} =\dot{\mathbf{s}}_0 + \alpha \mathbf{s}' \times (\mathbf{v}_n - \dot{\mathbf{s}}_0)
- \alpha' \mathbf{s}' \times [\mathbf{s}' \times (\mathbf{v}_n - \dot{\mathbf{s}}_0)],
\label{sdotmf}
\end{equation}
where $\alpha$ and $\alpha'$ are temperature-dependent friction coefficients \cite{Schwarz85}, and $\dot{\mathbf{s}}_0$ is calculated from Eq. (\ref{s0dot}). 

The numerical simulation method based on this model is described in detail elsewhere \cite{Schwarz85, Schwarz88,Tsubota00}. A vortex filament is represented by a single string of points separated by a distance $\Delta\xi$. The vortex configuration at a given time determines the velocity field in the fluid, thus moving the vortex filaments according to Eqs. (\ref{s0dot}) and (\ref{sdotmf}). 
Vortex reconnection should be properly included when simulating vortex dynamics.  A numerical study of a classical fluid shows that the close interaction of two vortices leads to their reconnection, chiefly because of the viscous diffusion of the vorticity \cite{Boratav92}. Schwarz assumed that two vortex filaments reconnect when they come within a critical distance of one another and showed that statistical quantities such as vortex line density were not sensitive to how the reconnections occur \cite{Schwarz85, Schwarz88}.  Even after Schwarz's study, it remained unclear as to whether quantized vortices can actually reconnect. However, Koplik and Levine solved directly the GP equation to show that two closely quantized vortices reconnect even in an inviscid superfluid \cite{Koplik93}. More recent simulations have shown that reconnections are accompanied by emissions of sound waves having wavelengths on the order of the healing length \cite{Leadbeater01, Ogawa02}. 

Starting with several remnant vortices under thermal counterflow, Schwarz studied numerically how these vortices developed into a statistical steady vortex tangle \cite{Schwarz88}.  The tangle was self-sustained by the competition between the excitation due to the applied flow and the dissipation through the mutual friction. The numerical results were quantitatively consistent with typical experimental results. This was a significant accomplishment for numerical research.

Here, we shall introduce some quantities that are characteristic of a vortex tangle. The line length density $L$ is defined as the total length of vortex cores in a unit volume. The mean spacing $\ell$ between vortices is given by $\ell=L^{-1/2}$. 

\subsection{The Gross-Pitaevskii (GP) model}
In a weakly interacting Bose system, the macroscopic wave function $\Psi(\mathbf{x},t)$ appears as the order parameter of Bose--Einstein condensation, obeying the Gross--Pitaevskii (GP) equation \cite{Pethick02}: 
\begin{equation}
i \hbar \frac{\partial \Psi(\mathbf{x},t)}{\partial t} = \biggl( - \frac{\hbar ^2}{2m}\nabla^2
+g |\Psi(\mathbf{x},t)|^{2}-\mu \biggr) \Psi(\mathbf{x},t).
\label{gpeq}
\end{equation}
Here, $g=4\pi \hbar^2 m/a$ represents the strength of the interaction characterized by the s-wave scattering length $a $, $m$ is the mass of each particle, and $\mu$ is the chemical potential. Writing $\Psi = | \Psi | \exp (i \theta)$, the squared amplitude $|\Psi|^2$ is the condensate density and the gradient of the phase $\theta$ gives the superfluid velocity  $\mathbf{v}_s = (\hbar/m) \nabla \theta$, which is a frictionless flow of the condensate. This relation causes quantized vortices to appear with quantized circulation. The only characteristic scale of the GP model is the coherence length defined by $\xi=\hbar/(\sqrt{2mg}| \Psi |)$, which gives the vortex core size. 
The GP model can explain not only the vortex dynamics but also phenomena related to vortex cores, such as reconnection and nucleation. However, the GP equation is not applicable quantitatively to superfluid $^4$He, which is not a weakly interacting Bose system. It is, however, applicable to Bose--Einstein condensation of a dilute atomic Bose gas \cite{Pethick02}.
 
\section{Modern research trends in QT}
Most older experimental studies on QT were devoted to thermal counterflow. Since this flow has no classical analogue, these studies do not contribute greatly to the understanding of the relationship between CT and QT. Since the mid 90s, important experimental studies on QT that did not focus on thermal counterflow have been published.
\subsection{New experiments on energy spectra}
The first important contribution was made by
Maurer and Tabeling \cite{Maurer98}, who confirmed experimentally the Kolmogorov spectrum in superfluid $^4$He for the first time.
A turbulent flow was produced in a cylinder by driving two counter-rotating disks.
The authors observed the local pressure fluctuations to obtain the energy spectrum.  
The experiments were conducted at three different temperatures 2.3 K, 2.08 K, and 1.4 K.
Both above and below the $\lambda$ point, the Kolmogorov spectrum was confirmed.
The observed behavior above the $\lambda$ point is not surprising because the system is a 
classical viscous fluid. 
However, it is not trivial to understand the Kolmogorov spectrum at two different temperatures below the $\lambda$ point.

The next significant step was a series of experiments on grid turbulence performed for superfluid $^4$He above 1 K by the Oregon group \cite{Smith93, Stalp99, Skrbek00a,Skrbek00b,Stalp02}.  
The flow through a grid is usually used to generate turbulence in classical fluid dynamics \cite{Frisch}. At a sufficient distance behind the grid, the flow displays a form of homogeneous isotropic turbulence. This method has also been applied to superfluid helium.  
In the experiments by the Oregon group, the helium was contained in a channel with a square cross section, through which a grid was pulled at a constant velocity. 
A pair of second-sound transducers was set into the walls of the channel.
When a vortex tangle appeared in a channel, it was detected by second-sound attenuation \cite{Vinen02}. The decay of the vorticity of the tangle created behind the towed grid was observed by the pair of transducers. In combining the observations with the decay of the turbulence, the authors made some assumptions. 
In fully developed turbulence, the energy dissipation rate can be shown to be given as 
$\epsilon=\nu \langle \omega^2 \rangle$,
where $\langle \omega^2 \rangle$ is the mean square vorticity (rot $\mathbf{v}$) in the flow \cite{Hinze75}.
The authors assumed that a similar formula applies in superfluid helium above 1 K.
They noted that the quantity $\kappa^2 L^2$ would be a measure of the mean square vorticity in the superfluid component. Hence, they assumed that in grid turbulence the dissipation rate is given by
$\epsilon=\nu' \kappa^2 L^2 $
with an effective kinematic viscosity $\nu'$. 
In order to combine this representation with the observations of second-sound attenuation for grid turbulence, the authors furthermore assumed that a quasi-classical flow appears at length scales much greater than $\ell$.  The flow is thought to come from a mechanism coupling the superfluid and the normal fluid by mutual friction, causing the fluid to behave like a one-component fluid \cite{Vinen00}. By choosing suitable values of $\nu'$ as a function of temperature \cite{Stalp02}, it was found that $\kappa L$ decays as $t^{-3/2}$.  

This characteristic decay $L \propto t^{-3/2}$ is quite important, because it is related to the Kolmogorov spectrum.  
Thus, we herein present the simple argument given in a previous review article \cite{Vinen02}. 
We assume that a turbulent fluid obeys the Kolmogorov spectrum of Eq. (\ref{eq-Kolmogorov}) in the inertial range of $D^{-1}< k< k_d$ with $D^{-1}<< k_d$.
The total energy is approximately given by
\begin{equation}
E=\int_{D^{-1}}^{k_d} C\varepsilon^{2/3}k^{-5/3}\approx \frac32 C \varepsilon^{2/3}D^{2/3}.
\end{equation}
If the turbulence decays slowly with time and the dissipation rate $\varepsilon$ is assumed to be time-dependent, we can write
\begin{equation}
\varepsilon=-\frac{dE}{dt}=-C\varepsilon^{-1/3}D^{2/3}\frac{d\varepsilon}{dt}.
\end{equation}
The solution gives the time-dependence of $\varepsilon$:
\begin{equation}
\varepsilon=27C^3D^2(t+t_0)^{-3},
\label{varepsilon}
\end{equation}
where $t_0$ is a constant. Combining Eq. (\ref{varepsilon}) with the above formula of $\varepsilon$ gives the decay of $L$:
\begin{equation}
L=\frac{(3C)^{3/2}D}{\kappa \nu'^{1/2}} (t+t_0)^{-3/2}.
\label{L-decay}
\end{equation}
This behavior has been observed, and a quantitative comparison with observations at any temperature gives $\nu'$ as a function of temperature. 
The observation of the decay $L \sim t^{-3/2}$ indicates that the Kolmogorov spectrum applies in turbulence, although it is not necessarily direct proof. 
Note that this simple analysis is applicable only when the maximum length scale of the turbulent energy saturates at the size of the channel. For the complete dynamics, a more complicated decay of vortices has been observed and has been found to be consistent with the classical model of the Kolmogorov spectrum \cite{Stalp99, Skrbek00a, Skrbek00b}. This type of decay is also observed at very low temperatures in the turbulence induced by an impulsive spin down for superfluid $^4$He \cite{Walmsley07} and in grid turbulence for superfluid $^3$He-B \cite{Bradley06, PickettPLTP}.   

\subsection{Energy spectra at finite temperatures}
These observations lead us to inquire as to the nature of the velocity field that gives rise to the observed energy spectrum. Vinen considered the situation theoretically \cite{Vinen00} and proposed that it is likely that the superfluid and normal fluid are coupled by mutual friction at scales larger than the characteristic scale $\ell$  of the vortex tangle. If so, the two fluids behave as a one-component fluid at these scales, where mutual friction does not cause dissipation. Since the normal fluid is viscous, the two coupled fluids can be turbulent and obey the Kolmogorov spectrum. The observation of the energy spectrum by Maurer et al. and Stalp et al. should support the idea of the coupled dynamics of two fluids. This behavior has been confirmed numerically as well\cite{Kivotides07}. At small scales, the two fluids should be decoupled, so that both mutual friction and normal fluid viscosity operate.  What then happens to the energy spectrum? Although theoretical consideration has been given to this problem \cite{L'vov06}, the answer remains controversial and has not yet been clarified.  While this is an important problem, it is not investigated in the present study. 

\subsection{Modern research trends}
The following three trends are currently the primary research areas in QT. The first is the energy spectra and the dissipation mechanism at zero temperature \cite{TsubotaPLTP}.  The second is QT created by vibrating structures \cite{VinenPLTP}. The third is visualization of QT \cite{SciverPLTP}. The remainder of this article is devoted to the review of these topics.

\section{Energy spectra and dissipation at zero temperature}
What happens to QT at zero temperature is not so trivial.  The first problem is determining the nature of the energy spectrum of turbulence for the pure superfluid component \cite{TsubotaPLTP}. If QT has a classical analogue, the energy spectrum is expected to obey the Kolmogorov law and have an inertial range in which the energy is transferred self-similarly from large to small scales. In QT at zero temperature, any rotational motion should be carried by quantized vortices. Since quantized vortices are definite topological defects, the cascade can be attributed directly to their dynamics, which is different from the case for CT.  The second problem is determining how the energy is transferred from large to small scales. There is no dissipative mechanism at large scales. However, some dissipative mechanism should operate at small scales, as described below. A vortex tangle has a characteristic scale $\ell$, which is defined by the mean spacing between vortex lines. At scales greater than $\ell$, a Richardson cascade should transfer the energy through the breakup of vortices. However, the Richardson cascade becomes ineffective at small scales, especially below $\ell$. What mechanism cascades the energy instead of the Richardson cascade at these scales?  
The third problem is understanding the dissipation in the system.  The first possibility is acoustic emission at vortex reconnections. In classical fluid dynamics, vortex reconnections cause acoustic emission. In quantum fluids, numerical simulations of the GP model shows acoustic emission at every reconnection event \cite{Leadbeater01}. However, this mechanism is thought to be unimportant because of the very short coherence length.
The second possible mechanism is the radiation of sound (phonons) by the oscillatory motion of vortex cores.  We will return to these mechanisms, as well as other possibilities.

\subsection{Energy spectra}
No experimental studies have addressed this issue directly, although a few numerical studies have been conducted. The first study was performed by Nore et al. using the GP model \cite{Nore97a, Nore97b}. They solved the GP equation numerically starting from Taylor--Green vortices, and followed the time development. The quantized vortices become tangled and the energy spectra of the incompressible kinetic energy seemed to obey the Kolmogorov law for a short period, although the energy spectra eventually deviated from the Kolmogorov law.  The second study was performed by the vortex filament model\cite{Araki02}, and the third study was performed by the modified GP model \cite{Kobayashi05a, Kobayashi05b}.   
\subsubsection{Energy spectra by the vortex filament model}
Using the vortex filament model, Araki et al. generated a vortex tangle arising from Taylor--Green vortices and obtained an energy spectrum consistent with the Kolmogorov law\cite{Araki02}.  It would be informative to describe how to obtain the energy spectra under the vortex filament model. 
The energy spectrum is calculated by Fourier transform of the superfluid velocity $\mathbf{v}_s(\mathbf{r})$, which is determined by the configuration of quantized vortices. Thus, the energy spectrum can be calculated directly from the configuration of the vortices. Using the Fourier transform $\mathbf{v}_s(\mathbf{k})=(2\pi)^{-3} \int d\mathbf{r} \mathbf{v}_s(\mathbf{r})\exp (-i \mathbf{k} \cdot \mathbf{r}) $ and Parseval's theorem $\int d\mathbf{k} |\mathbf{v}_s(\mathbf{k})|^2=(2\pi)^{-3} \int d\mathbf{r} |\mathbf{v}_s(\mathbf{r})|^2$, the kinetic energy of the superfluid velocity per unit mass is
 \begin{equation}
E=\frac12 \int d\mathbf{r} |\mathbf{v}_s(\mathbf{r})|^2=\frac{(2\pi)^3}{2} \int d\mathbf{k} |\mathbf{v}_s(\mathbf{k})|^2.
\end{equation}
The vorticity $\mathbf{\omega}(\mathbf{r})={\rm rot} \mathbf{v}_s(\mathbf{r})$ is represented in Fourier space as $\mathbf{v}_s(\mathbf{k})=i \mathbf{k} \times \mathbf{\omega}(\mathbf{k}) /|\mathbf{k}|^2$, so that we have $E=\left( (2\pi)^3/2 \right) \int d\mathbf{k} |\mathbf{\omega}(\mathbf{k})|^2/|\mathbf{k}|^2$. The vorticity is concentrated on the vortex filament, represented by
\begin{equation} \mathbf{\omega}(\mathbf{r})=\kappa \int d\xi \mathbf{s}'(\xi) \delta (\mathbf{s}(\xi)-\mathbf{r}), \end{equation} 
which can be rewritten as \begin{equation} \mathbf{\omega}(\mathbf{k})=\frac{\kappa}{(2\pi)^2}  \int d\xi \mathbf{s}'(\xi) \exp (-i \mathbf{s}(\xi) \cdot\mathbf{k}). \end{equation}  Using the definition of the energy spectrum $E(k)$ from $E= \int_0^{\infty} dk E(k)$, these relations yield
\begin{eqnarray}
E(k)=\frac{\kappa^2}{2(2\pi)^3} \int \frac{d\Omega_k}{|\mathbf{k}|^2} \int\int d\xi_1
d\xi_2 \mathbf{s}'(\xi_1)\cdot \mathbf{s}'(\xi_2)  \nonumber \\  \times \exp (-i \mathbf{k} \cdot(\mathbf{s}(\xi_1)-\mathbf{s}(\xi_2))),
\label{spectrum}
\end{eqnarray}
where $d\Omega_k=k^2 \sin \theta_k d\theta_k d\phi_k$ is the volume element in spherical coordinates. 

Starting from a Taylor--Green vortex and following the vortex motion under the full Biot--Savart law without mutual friction, Araki et al. obtained a roughly homogeneous and isotropic vortex tangle \cite{Araki02}. This was a decaying turbulence, dissipated by the cutoff of the smallest vortices, the size of which is comparable to the space resolution. Initially, the energy spectrum has a large peak at the largest scale, where the energy is concentrated. The spectrum changes as the vortices become homogeneous and isotropic.  The energy spectrum of the vortex tangle at some late stage was quantitatively consistent with the Kolmogorov spectrum in the small $k$ region. By monitoring the development of the vortex size distribution, the decay of a tangle is found to be sustained by a Richardson cascade process. These results support the quasi-classical picture of the inertial range in QT at very low temperatures.
\subsubsection{Energy spectra by the GP model}
As the third trial, the Kolmogorov spectra was confirmed for both decaying \cite{Kobayashi05a} and steady \cite{Kobayashi05b} QT by the modified GP model.  The normalized GP equation is
\begin{equation} i\frac{\partial}{\partial t}\Phi(\mathbf{x},t)=[-\nabla^2-\mu+g|\Phi(\mathbf{x},t)|^2]\Phi(\mathbf{x},t),\label{eq-GP} \end{equation}
which determines the dynamics of the macroscopic wave function 
$\Phi(\mathbf{x},t)=f(\mathbf{x},t)\exp[ i \phi(\mathbf{x},t)]$. The condensate density is $|\Phi(\mathbf{x},t)|^2=f(\mathbf{x},t)^2$, and the superfluid velocity $\mathbf{v}(\mathbf{x},t)$ is given by $\mathbf{v}(\mathbf{x},t)=2\nabla\phi(\mathbf{x},t)$. The vorticity $\omega(\mathbf{x},t)={\rm rot} \mathbf{v}(\mathbf{x},t)$ vanishes everywhere in a single-connected region of the fluid and thus all rotational flow is carried by quantized vortices. In the core of each vortex, $\Phi(\mathbf{x},t)$ vanishes so that the circulation around the core is quantized by $4\pi$. The vortex core size is given by the healing length $\xi=1/f\sqrt{g}$.

Note that the hydrodynamics described by the GP model is compressible, which is different from the case of the vortex filament model. The total energy of the GP model
\begin{equation} E(t)=\frac{1}{N} \int d\mathbf{x}\:\Phi^\ast(\mathbf{x},t)\Big[-\nabla^2+\frac{g}{2}f(\mathbf{x},t)^2\Big]\Phi(\mathbf{x},t),\end{equation}
is represented by the sum of the interaction energy $E_{int}(t)$, the quantum energy $E_{q}(t)$, and the kinetic energy $E_{kin}(t)$ \cite{Nore97a, Nore97b}:
\begin{eqnarray} E_{int}(t)=\frac{g}{2N}\int d\mathbf{x}\:f(\mathbf{x},t)^4,  E_{q}(t)=\frac{1}{N}\int d\mathbf{x}\:[\nabla f(\mathbf{x},t)]^2,\nonumber\\ E_{kin}(t)=\frac{1}{N}\int d\mathbf{x}\:[f(\mathbf{x},t)\nabla\phi(\mathbf{x},t)]^2.\label{eq-energy-definition} \end{eqnarray}
The kinetic energy is furthermore divided into a compressible part $E_{kin}^{c}(t)$ due to compressible excitations and an incompressible part $E_{kin}^{i}(t)$ due to vortices.
If the Kolmogorov spectrum is observed, the spectrum should be that for the incompressible kinetic energy $E_{kin}^{i}(t)$.

The failure to obtain the Kolmogorov law under the pure GP model \cite{Nore97a, Nore97b} would be attributable to the following reasons. Note that the situation here is decaying turbulence. Although the total energy $E(t)$ was conserved, $E_{kin}^{i}(t)$ decreased with increasing $E_{kin}^{c}(t)$. This was because many compressible excitations were created through repeated vortex reconnections \cite{Leadbeater01, Ogawa02} and disturbed the Richardson cascade of quantized vortices even at large scales. 

Kobayashi and Tsubota overcame the difficulties of Nore et al. and obtained the Kolmogorov spectra in QT and clearly revealed the energy cascade \cite{Kobayashi05a, Kobayashi05b}.  They performed numerical calculation for the Fourier transformed GP equation with dissipation:
\begin{eqnarray}(i-\tilde{\gamma}(\mathbf{k}))\frac{\partial}{\partial t}\tilde{\Phi}(\mathbf{k},t)=[k^2-\mu(t)]\tilde{\Phi}(\mathbf{k},t) \nonumber\\ +\frac{g}{V^2}\sum_{\mathbf{k}_1,\mathbf{k}_2}\tilde{\Phi}(\mathbf{k}_1,t)\tilde{\Phi}^\ast(\mathbf{k}_2,t) \times\tilde{\Phi}(\mathbf{k}-\mathbf{k}_1+\mathbf{k}_2,t). \label{eq-Fourier-GP1} \end{eqnarray}    
Here, $\tilde{\Phi}(\mathbf{k},t)$ is the spatial Fourier component of $\Phi(\mathbf{x},t)$ and $V$ is the system volume. The healing length is given by $\xi=1/|\Phi|\sqrt{g}$.  The dissipation should have the form $\tilde{\gamma}(\mathbf{k})=\gamma_0\theta(k-2\pi/\xi)$ with the step function $\theta$, which dissipates only the excitations smaller than $\xi$. This form of dissipation can be justified by the coupled analysis of the GP equation for the macroscopic wave function and the Bogoliubov--de Gennes equations for thermal excitations \cite{Kobayashi06a}. 

First, Kobayashi et al. confirmed the Kolmogorov spectra for decaying turbulence \cite{Kobayashi05a}.  To obtain a turbulent state, they started the calculation from an initial configuration in which the density was uniform and the phase of the wave function had a random spatial distribution. The initial wave function was dynamically unstable and soon developed into fully developed turbulence with many quantized vortex loops. The spectrum $E_{kin}^{i}(k,t)$ of the incompressible kinetic energy was then found to obey the Kolmogorov law. 

A more elaborate analysis of steady QT was performed by introducing energy injection at large scales as well as energy dissipation at small scales \cite{Kobayashi05b}. Energy injection at large scales was effected by moving a random potential $V(\mathbf{x},t)$. Numerically, Kobayashi et al. placed random numbers between $0$ and $V_0$ in space-time $(\mathbf{x},t)$ at intervals of $X_0$ for space and $T_0$ for time and connected them smoothly using a four-dimensional spline interpolation.  The moving random potential exhibited a Gaussian two-point correlation:
\begin{equation} \langle V(\mathbf{x},t)V(\mathbf{x}^{\prime},t^{\prime})\rangle=V_0^2\exp\Big[-\frac{(\mathbf{x}-\mathbf{x}^{\prime})^2}{2X_0^2}-\frac{(t-t^{\prime})^2}{2T_0^2}\Big].\label{eq-random-correlation} \end{equation}
This moving random potential had a characteristic spatial scale of $X_0$. Small vortex loops were first nucleated by the random potential, growing to the scale of $X_0$ by its motion subjected to Eq. ({\ref{eq-random-correlation}}). The vortex loops were then cast into the Richardson cascade.  If steady QT is obtained by the balance between the energy injection and the dissipation, it should have an energy-containing range of $k<2\pi /X_0$, an inertial range of $2\pi/X_0<k<2\pi/\xi$, and an energy-dissipative range of $2\pi/\xi <k$.

\begin{figure}[htb] \centering \begin{minipage}{0.49\linewidth} \begin{center} \includegraphics[width=.99\linewidth]{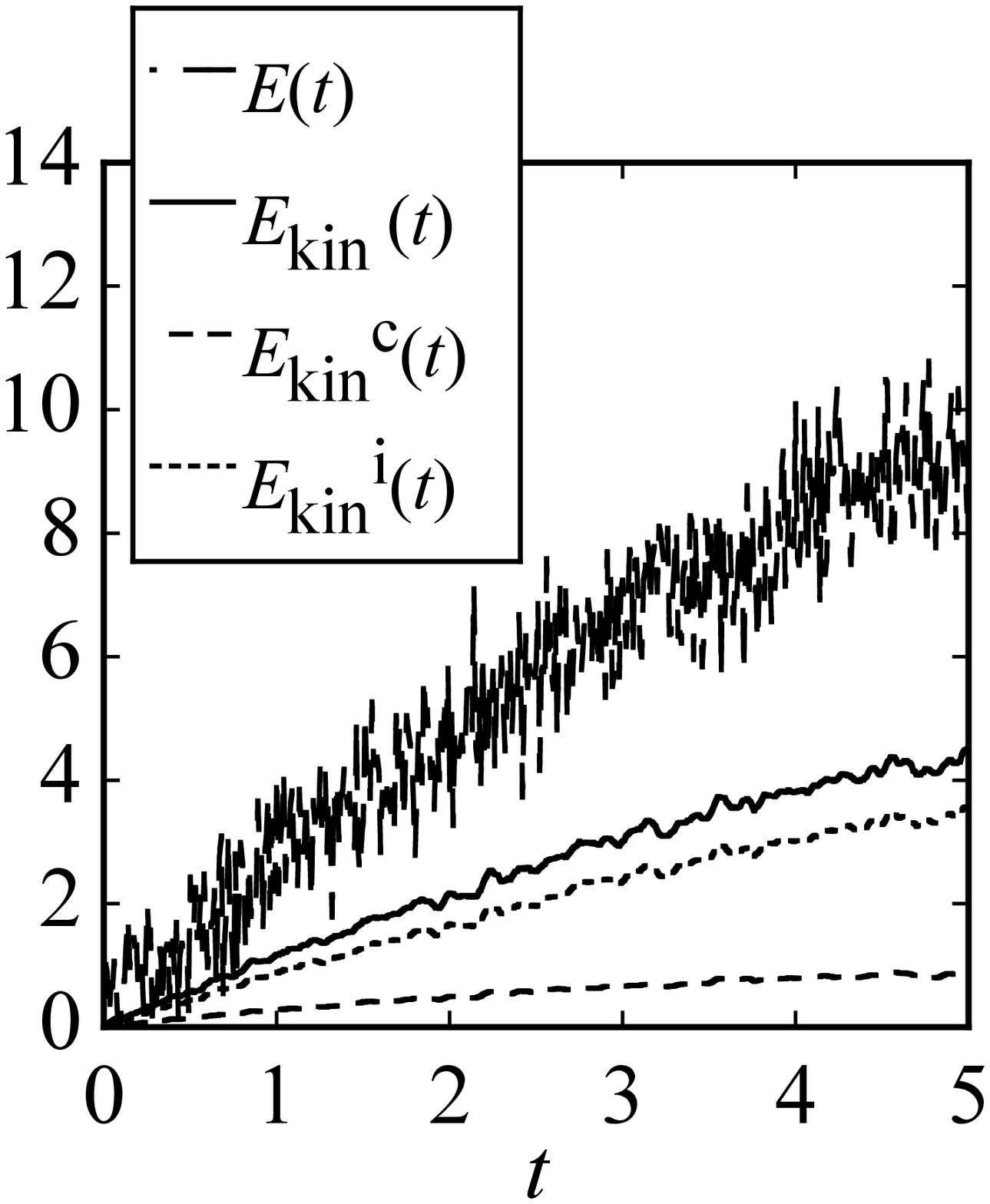}\\ (a) \end{center} \end{minipage} \begin{minipage}{0.49\linewidth} \begin{center} \includegraphics[width=.99\linewidth]{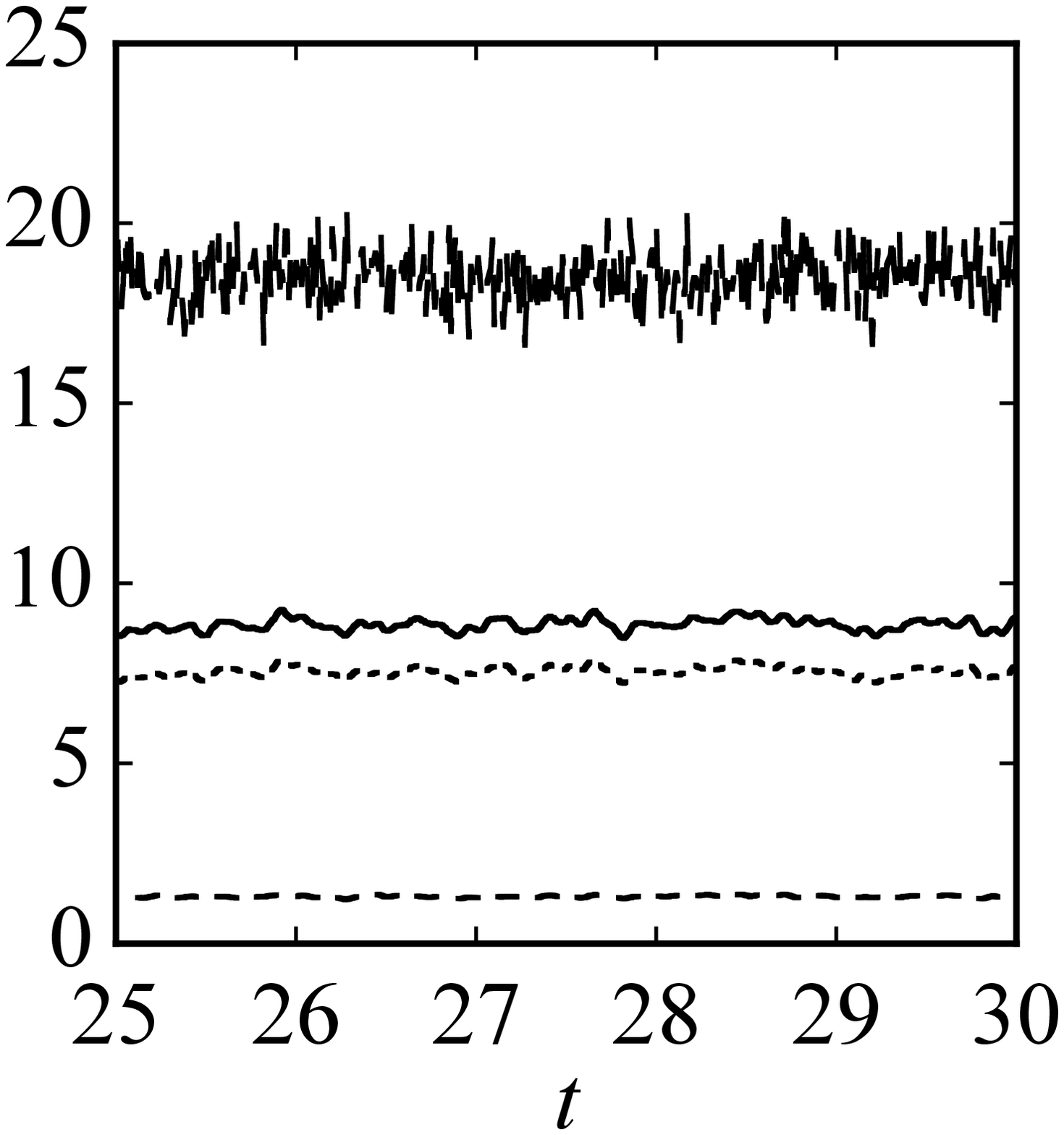}\\ (b) \end{center} \end{minipage}
\caption{Time development of $E(t)$, $E_{kin}(t)$, $E_{kin}^{c}(t)$, and $E_{kin}^{i}(t)$ at (a) the initial stage $0\le t\le 5$ and (b) a later stage $25\le t\le 30$ \cite{Kobayashi05b}.} \label{fig-steady-energy} \end{figure}
\begin{figure}[htb] \centering \begin{minipage}[t]{0.95\linewidth} \begin{center} \includegraphics[width=.99\linewidth]{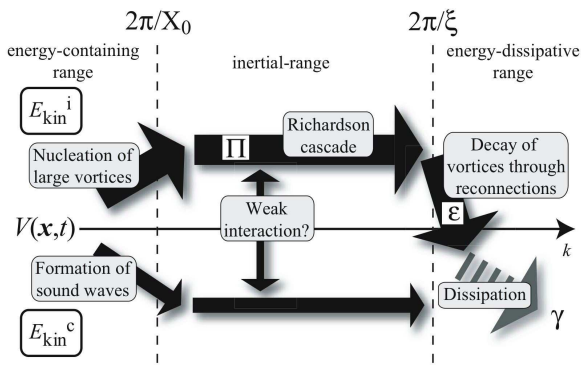} \end{center} \end{minipage} 
\caption{ Flow of the incompressible kinetic energy $E_{kin}^{i}(t)$ (upper half of diagram) and compressible kinetic energy $E_{kin}^{c}(t)$ (lower half) in wavenumber space \cite{Kobayashi05b}.} \label{fig-energy-current}\end{figure}
A typical simulation of steady turbulence was performed for $V_0=50$, $X_0=4$, and $T_0=6.4\times10^{-2}$. The dynamics started from the uniform wave function. Figure \ref{fig-steady-energy} shows the time development of each energy component. The moving random potential nucleates sound waves as well as vortices, but both figures show that the incompressible kinetic energy $E_{kin}^{i}(t)$ due to vortices is dominant in the total kinetic energy $E_{kin}(t)$.  The four energies are almost constant for $t\gtrsim 25$, and steady QT was obtained. 

Such a steady QT enables us to investigate the energy cascade. Here, we expect an energy flow in wavenumber space similar to that in Fig. \ref{fig-energy-current}. The upper half of the diagram shows the kinetic energy $E_{kin}^{i}(t)$ of quantized vortices, and the lower half shows the kinetic energy $E_{kin}^{c}(t)$ of compressible excitations. In the energy-containing range $k<2\pi /X_0$, the system receives incompressible kinetic energy from the moving random potential. During the Richardson cascade process of quantized vortices, the energy flows from small to large $k$ in the inertial range $2\pi/X_0<k<2\pi/\xi$. In the energy-dissipative range $2\pi/\xi <k$, the incompressible kinetic energy transforms to compressible kinetic energy through reconnections of vortices or the disappearance of small vortex loops. The moving random potential also creates long-wavelength compressible sound waves, which are another source of compressible kinetic energy and also produce an interaction with vortices. However, the effect of sound waves is weak because $E_{kin}^{i}(t)$ is much larger than $E_{kin}^{c}(t)$, as shown in Fig. \ref{fig-steady-energy}. 

\begin{figure}[htb] \centering \begin{minipage}{0.49\linewidth} \begin{center} \includegraphics[width=.99\linewidth]{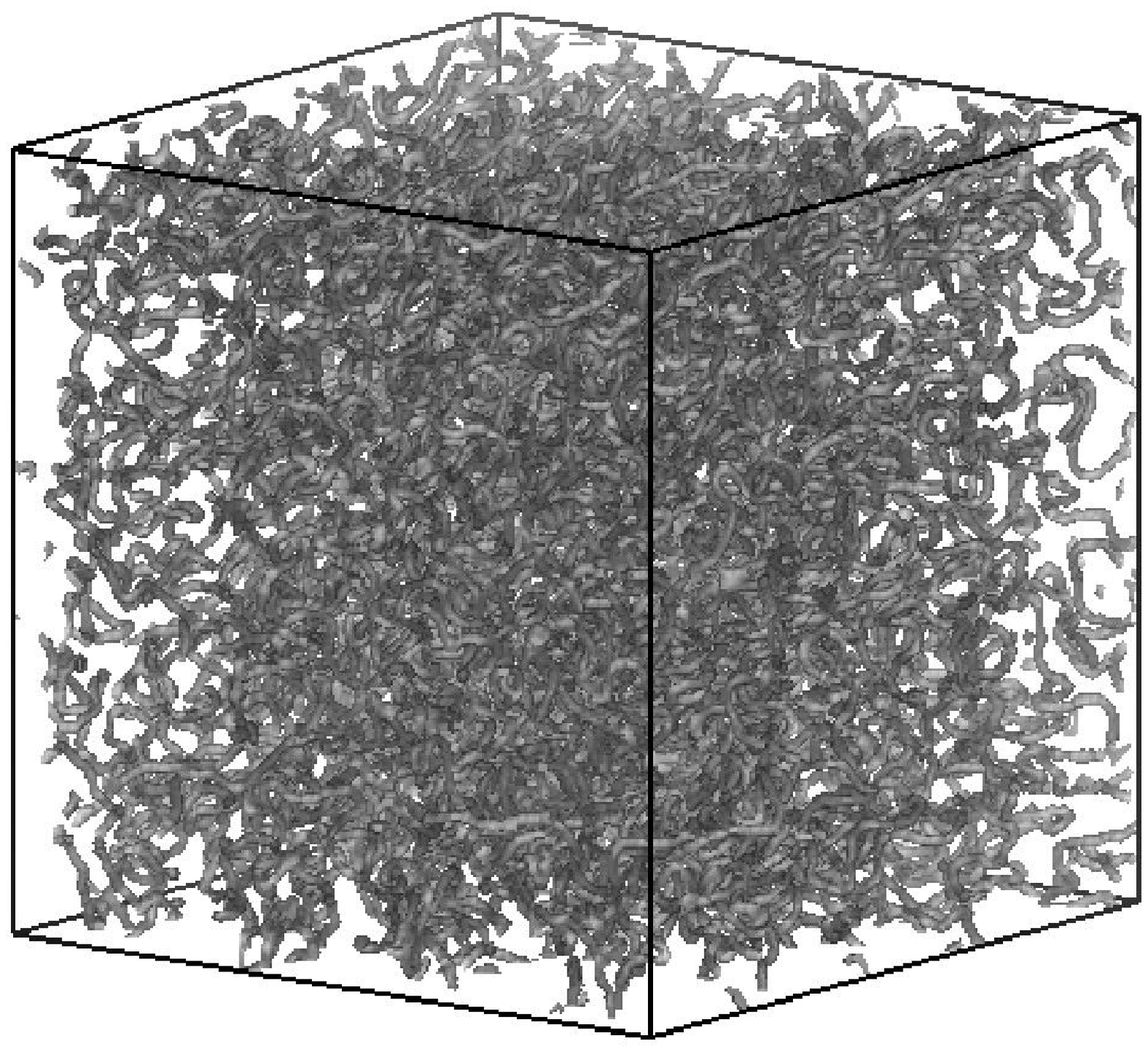}\\ (a) \end{center} \end{minipage} \begin{minipage}{0.45\linewidth} \begin{center} \includegraphics[width=.99\linewidth]{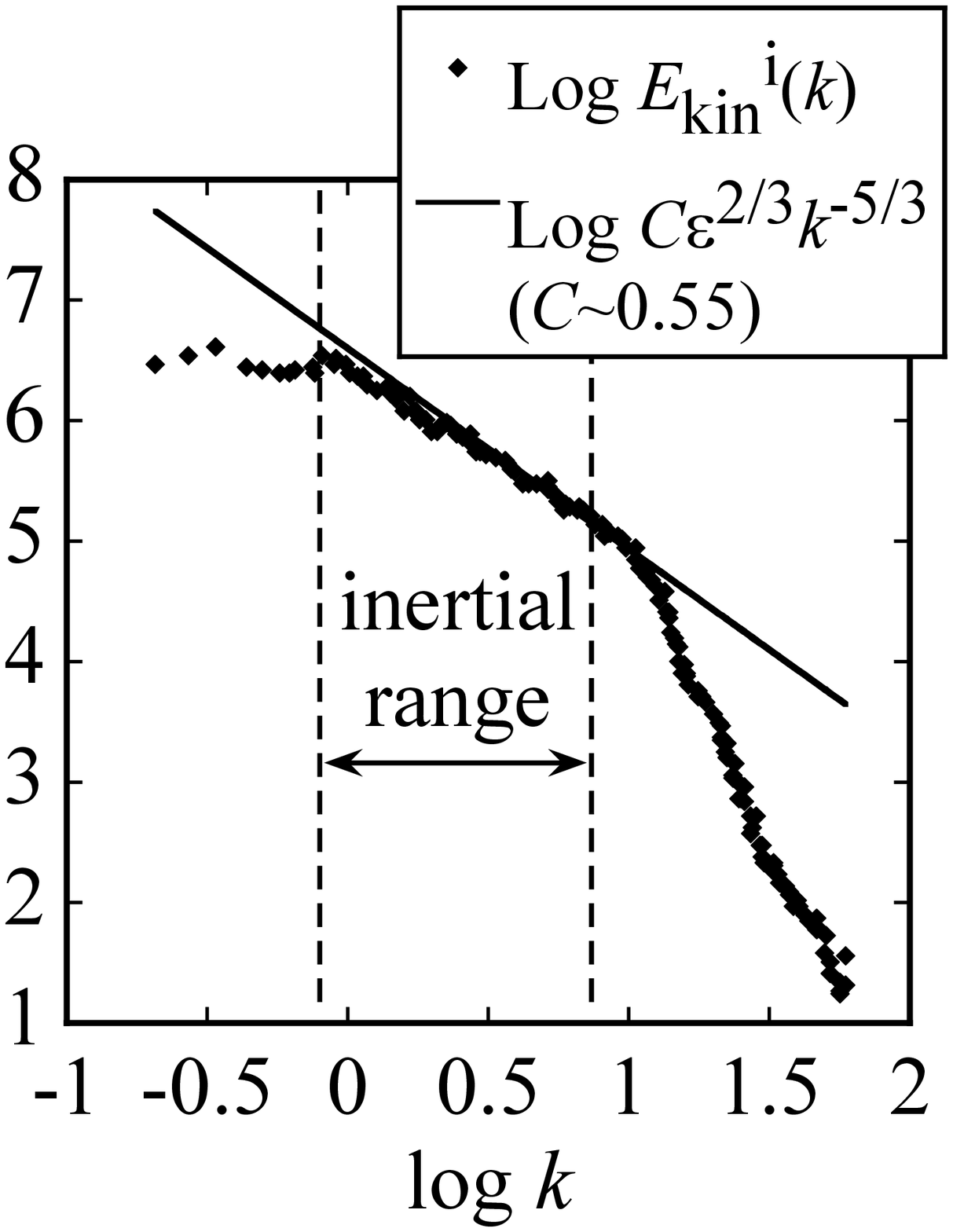}\\ (b) \end{center} \end{minipage}
\caption{(a) A typical vortex tangle. (b) Energy spectrum $E_{kin}^{i}(k,t)$ for QT. The plotted points are from an ensemble average of 50 randomly selected states at $t>25$. The solid line is the Kolmogorov law \cite{Kobayashi05b}.} \label{fig-steady-Kolmogorov} \end{figure}

This cascade can be confirmed quantitatively by checking whether the energy dissipation rate $\varepsilon$ of $E_{kin}^{i}(t)$ is comparable to the flux of energy $\Pi(k,t)$ through the Richardson cascade in the inertial range. Although the details are described in Reference 42, $\Pi(k,t)$ is found to be approximately independent of $k$ and comparable to $\varepsilon$.  As shown in Fig. \ref{fig-steady-Kolmogorov}, the energy spectrum is quantitatively consistent with the Kolmogorov law in the inertial range $2\pi/X_0<k<2\pi/\xi$ (b), which is equivalent to $0.79\lesssim k\lesssim 6.3$.

Kobayashi et al. studied the decay of QT under the same formulation \cite{Kobayashi06b}. After obtaining a steady tangle, they switched off the motion of the random potential and found that $L$ decayed as $t^{-3/2}$. 

\subsection{The Kelvin-wave cascade}
The arguments in the last section were chiefly limited to the large scale, usually larger than the mean spacing $\ell$ of a vortex tangle, in which the Richardson cascade is effective for transferring the energy from large to small scales. Here, we should consider what happens at smaller scales, for which the Richardson cascade should be less effective.  The most probable scenario is the Kelvin-wave cascade.  
A Kelvin-wave is a deformation of a vortex line into a helix with the deformation propagating as a wave along the vortex line \cite{Thomson80}. Kelvin-waves were first observed by making torsional oscillations in uniformly rotating superfluid $^4$He \cite{Hall58, Hall60}. The approximate dispersion relation for a rectilinear vortex is 
$\omega_k = (\kappa k^2)/(4\pi) (\ln(1/ka_0)+c)$ 
with a constant $c \sim 1$. Note that this $k$ is the wavenumber of an excited Kelvin-wave and is different from the wavenumber used for the energy spectrum in the last subsection.  At a finite temperature, a significant fraction of normal fluid damps Kelvin-waves through mutual friction.  At very low temperatures, however, mutual friction does not occur, and the only possible mechanism of dissipation is the radiation of phonons \cite{Vinen01}. Phonon radiation becomes effective only when the frequency becomes very high, typically on the order of GHz ($k \sim  10^{-1}$ nm$^{-1}$), so a mechanism is required to transfer the energy to such high wavenumbers in order for Kelvin-waves to be damped.  An early numerical simulation based on the vortex filament model showed that Kelvin-waves are unstable to the buildup of side bands \cite{Samuels90}. This indicates the possibility that nonlinear interactions between different Kelvin-wave numbers can transfer energy from small to large wavenumbers, namely the Kelvin-wave cascade.  This idea was first suggested by Svistunov \cite{Svistunov95} and was later developed and confirmed through theoretical and numerical analyses by Vinen, Tsubota and Mitani \cite{Vinen03}, and Kozik and Svistunov \cite{Kozik04, Kozik05a, Kozik05b}. 

Observation of the Kelvin-wave cascade is crucial.  Such studies are not easy for a vortex tangle. The easiest approach would be to consider rotation.  In a rotating vessel, quantized vortices form a vortex lattice parallel to the rotational axis.  By oscillating the vortices, Kelvin-waves can be excited. This approach was used in the pioneering experiments of Hall \cite{Hall58, Hall60}. The challenge is detecting the Kelvin-wave cascade when it occurs. There are two possible methods. The first is the direct visualization of vortices. Recently, remarkable progress has been made in visualizing superflow and quantized vortices \cite{Zhang05, Bewley06}, as described in Section 6.4.  Bewley et al. visualized quantized vortices by trapping micron-sized solid hydrogen particles \cite{Bewley06}. They also observed a vortex array under rotation.  Applying this technique to the Kelvin-wave cascade could reveal important information directly. The second method is to observe acoustic emission resulting from a Kelvin-wave cascade. Since the frequency of emitted phonons is estimated to be of GHz order, this observation may not be easy and presents a challenging experimental problem.

\subsection{Classical-quantum crossover}
An important trend is investigation of the nature of the transition between Richardson (classical) and Kelvin-wave (quantum) cascades. Several theoretical considerations on this topic have been investigated \cite{L'vov07, Kozik08}, although few numerical or experimental studies have been conducted.  

L'vov et al. theoretically discussed a bottleneck crossover between the two regions \cite{L'vov07}. Their investigation was based on the idea that a vortex tangle obeying the Kolmogorov spectrum is polarized to some degree, and they viewed the vortex tangle as a set of vortex bundles.  When they tried to connect the energy spectra between two cascades, a serious mismatch occurred at the crossover scale between the classical and quantum spectra. This mismatch prevented the energy flux from propagating fully through the crossover region, which is referred to as the bottleneck effect.  In order to remedy this mismatch, L'vov et al. used warm cascade solutions \cite{Cannaughton04} and proposed a thermal-equilibrium type spectrum between the classical and quantum spectra. The analysis of L'vov et al. was based on the assumption that the coarse-grained macroscopic picture of quantized vortices remains valid down to a scale of $\ell$. Without this assumption, Kozik and Svistunov theoretically investigated the details of the structure of the vortex bundle in the crossover region \cite{Kozik08}. Depending on the types of vortex reconnections, the crossover range near $\ell$ was further divided into three subranges, resolving the mismatch between two cascades. However, this topic is not yet fixed, and remains controversial.

\subsection{Possible dissipation mechanism at zero temperature}
Here, we should investigate the possible dissipation mechanisms of QT at very low temperatures, in which the normal fluid component is so negligible that mutual friction does not occur. In this case, there is no dissipation at relatively large scales, and dissipation can only occur at small scales. Energy at large scales should be transferred to smaller scales by a Richardson cascade and then a Kelvin-wave cascade until the dissipation becomes effective. The description presented herein is only schematic. Please refer to other papers for a more detailed discussion \cite{Vinen00, Vinen02}.
 
Dissipation of QT at zero temperature was first proposed by Feynman \cite{Feynman}.  Feynman suggested that a large vortex loop should be broken up into smaller loops through repeated vortex reconnections, which is essentially identical to the Richardson cascade. Feynman thought that the smallest vortex ring having a radius comparable to the atomic scale must be a roton, although this belief is not currently accepted universally. Later, Vinen considered the decay of superfluid turbulence in order to understand his experimental results for thermal counterflow \cite{Vinen57a, Vinen57b}, leading him to propose Vinen's equation \cite{Vinen57c}. Although the experiments were performed at finite temperatures, the physics is important in order to study the decay of QT, as described briefly herein. Vortices in turbulence are assumed to be approximately evenly spaced with a mean spacing of $\ell=L^{-1/2}$. The energy of the vortices then spreads from the vortices of wavenumber $1/\ell$ to a wide range of wavenumbers. The overall decay of the energy is governed by the characteristic velocity $v_s=\kappa/2 \pi \ell$ and the time constant $\ell/v_s$ of the vortices of size $\ell$,  giving 
\begin{equation}  \frac{dv_s^2}{dt}=-\chi \frac{v_s^2}{\ell/v_s}=-\chi \frac{v_s^3}{\ell}, \end{equation}
where $\chi$ is a dimensionless parameter that is generally dependent on temperature. Rewriting this equation using $L$, we obtain 
\begin{equation} \frac{dL}{dt}=-\chi \frac{\kappa}{2\pi}L^2. \label{Vinen eq}\end{equation}
This is called Vinen's equation. The original Vinen's equation also includes a term of vortex amplification, which is not shown here. This equation describes the decay of $L$, the solution of which is 
\begin{equation} \frac{1}{L}=\frac{1}{L_0}+\chi \frac{\kappa}{2\pi}t, \label{sol Vinen eq}\end{equation}
where $L_0$ is $L$ at $t=0$.  The thermal counterflow observations are well described by this solution, which gives the values of $\chi$ as a function of temperature \cite{Vinen57c}.  The decay at finite temperatures is due to mutual friction. The value of $\chi$ at zero temperature is obtained by a numerical simulation of the vortex filament model \cite{Tsubota00}. 

What causes the different types of decay of $L \sim t^{-3/2}$ of Eq. (\ref{L-decay}) and $L \sim t^{-1}$ of Eq. (\ref{sol Vinen eq})?   The different types of decay originate from the different structures of vortex tangles in QT.  When the energy spectrum of a vortex tangle obeys the Kolmogorov law, the tangle is self-similar in the inertial range, and most energy is in the largest vortex.   The line length density then decays as $L \sim t^{-3/2}$, as described in Section 4.1.  If a vortex tangle is random and has little correlation, however, the only length scale is $\ell = L^{-1/2}$, yielding a decay of $L \sim t^{-1}$, as described herein. Therefore, the observation of how $L$ decays is helpful in understanding the structure of vortex tangles in QT.

However, these studies do not explain what happens to QT at zero temperature at small scales. There are several possibilities.  The first is acoustic emission at vortex reconnections. In classical fluid dynamics it is known that vortex reconnections cause acoustic emission. In quantum fluids, numerical simulations of the GP model show that acoustic emission occurs at every reconnection \cite{Leadbeater01, Ogawa02}. The dissipated energy during each reconnection is approximately $3\xi$ times the vortex line energy per unit length.  We can estimate how $L$ decays by this mechanism. The number of reconnection events per volume per time is on the order of $\kappa L^{5/2}$ \cite{Tsubota00}. Since each reconnection should reduce the vortex length by the order of $\xi$, the decay of $L$ can be described as
$ dL/dt=-\chi_1 \xi \kappa L^{5/2}$
with a constant $\chi_1$ of order unity. The solution is 
\begin{equation}  \frac{1}{L^{3/2}}= \frac{1}{L_0^{3/2}}+\frac32 \chi_1 \xi \kappa t. \end{equation}
In superfluid helium, especially $^4$He, $\xi$ is so small that decay due to this mechanism should be negligible. 

The second possible mechanism is the radiation of sound (phonons) by the oscillatory motion of a vortex core.  The characteristic frequencies of vortex motion on a scale $\ell$ are of order $\kappa/\ell^2$, which is too small to cause effective radiation \cite{Vinen00}. In order to make the radiation effective, the vortices should form small-scale structures, which would be realized by two consecutive processes. The first process is due to vortex reconnections \cite{Svistunov95}. In a dense isotropic tangle, vortex reconnections occur repeatedly at a rate of order $\kappa/\ell^5$ per unit volume \cite{Tsubota00}. When two vortices approach one another, they twist so that they become locally antiparallel, and reconnect  \cite{Schwarz85}. They then separate leaving a local small-scale structure. However, even such local structures created by single reconnection events are insufficient to cause effective acoustic radiation. The next process occurs due to a Kelvin-wave cascade described in Section 5.2. The Kelvin-wave cascade could transfer the energy into the small scales in which the radiation of sound becomes effective.

In actual experiments, we may have to consider the effects of vortex diffusion, even though these effects are not due to dissipation. Using the vortex filament model, Tsubota et al. numerically studied how an inhomogeneous vortex tangle diffuses \cite{Tsubota03b}. The obtained diffusion of the line length density $L(\mathbf{x},t)$ was well described by
\begin{equation} \frac{dL(\mathbf{x},t)}{dt}=-\chi \frac{\kappa}{2\pi}L(\mathbf{x},t)^2+ D \nabla^2 L(\mathbf{x},t).\end{equation} 
Without the second term on the right-hand side, this is Vinen's equation (\ref{Vinen eq}) \cite{Vinen57c}. The second term represents the diffusion of a vortex tangle. The numerical results show that the diffusion constant $D$ is approximately 0.1$\kappa$.  The dimensional argument indicates that $D$ must be on the order of $\kappa$ \cite{Tsubota03b}.  

\section{ New QT experiments}
This section reviews briefly recent important experiments on QT, although additional research is  needed for all of these experiments.

\subsection{Temperature-dependent transition to QT}
Krusius et al. conducted a series of experimental studies on QT in superfluid $^3$He -B under rotation \cite{Finne06b,KrusiusPLTP,Finne03,Finne04b,Finne06a, Eltsov06b} and observed the strong temperature-dependent transition of a few seed vortices to turbulence.  A seed vortex can develop into turbulence when the temperature is lower than the onset temperature $T_{\text{on}}$, whereas the seed vortex does not lead to turbulence above $T_{\text{on}}$. The onset temperature $T_{\text{on}}$ is approximately 0.6$T_{\text{c}}$ with the superfluid transition temperature $T_{\text{c}}$ being independent of flow velocity\cite{Finne04b}. The key characteristics of using  $^3$He -B for research on QT are as follows. First, the normal fluid component is too viscous to become turbulent, which is very different from the case of superfluid $^4$He. Secondly, monitoring the NMR absorption spectra enables us to count the number of quantized vortices, which is not possible in $^4$He.   

\begin{figure}[htb] \centering \begin{minipage}[t]{0.95\linewidth} \begin{center} \includegraphics[width=.99\linewidth]{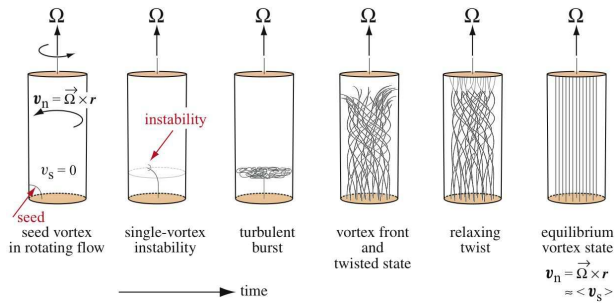} \end{center} \end{minipage} 
\caption{Schematic vortex instability and turbulence in rotating $^3$He-B in the turbulent temperature region studied by the Helsinki group \cite{KrusiusPLTP}.} \label{Helsinki}\end{figure}

Figure \ref{Helsinki} shows schematically the vortex instability and turbulence in rotating
3He-B in the turbulent temperature region studied by the Helsinki group.
A seed vortex is injected into $^3$He-B in a cylindrical vessel rotating with the angular velocity $\Omega$.  In the laboratory frame, the normal fluid component causes solid body rotation, while the superfluid component remains at rest. When the temperature is higher than $T_{\text{on}}$, the vortex just orientates along the rotational axis because of the mutual friction.  If the temperature is lower than $T_{\text{on}}$, however, the vortex becomes unstable, through several vortex reconnections that occur chiefly near the boundary\cite{Finne06a}, splitting into lots of vortices, and eventually reaching an equilibrium vortex lattice state. While the turbulent front propagates to the vortex-free region, the front takes on the beautiful "twisted vortex state" \cite{Eltsov06b}.  The onset temperature is determined by considering the dynamic mutual friction parameter $\zeta=(1-\alpha')/\alpha$, where $\alpha$ and $\alpha'$ are the mutual friction coefficients appearing in Eq. (\ref{sdotmf}) \cite{Finne06b}. This parameter $\zeta$ works in this system like the Reynolds number in a usual fluid.  


\subsection{Dissipation at very low temperatures}
Recently a few experimental studies on these topics have been conducted, showing the reduction of the dissipation at low temperatures. Eltsov et al. studied the vortex front propagation into a region of vortex-free flow in rotating superfluid $^3$He-B by NMR measurements following a series of investigations \cite{Eltsov07}. The observed front velocity as a function of temperature shows the transition from laminar through quasiclassical turbulent to quantum turbulent.  The front velocity is related to the effective dissipation, which exhibits a peculiar reduction at very low temperatures below approximately 0.25$T_c$ with the critical temperature $T_c$. Eltsov et al. claim that this is attributable to the bottleneck effect. Walmsley et al. made another observation of the effective viscosity of turbulence in superfluid $^4$He \cite{Walmsley07}.  Turbulence was produced by an impulsive spin down from an angular velocity to rest for a cube-shaped container, and the line length density was measured by scattering negative ions. The observed effective kinematic viscosity showed the striking reduction at low temperatures below approximately 0.8 K.  In this case, the bottleneck effect may not be so significant. The authors believe this may be due to another mechanism, namely, the difficulty in transferring energy through wavenumbers from the three-dimensional Richardson  cascade to the one-dimensional Kelvin-wave cascade. 

\subsection{QT created by vibrating structures}
Recently, vibrating structures, such as discs, spheres, grids, and wires, have been widely used for research into QT \cite{VinenPLTP}. Despite detailed differences between the used structures, the experiments show some surprisingly common phenomena. This trend started with the pioneering observation of QT on an oscillating microsphere by J\"ager et al.\cite{Jager95}. 

The sphere used by J\"ager et al. had a radius of approximately 100 $\mu$m. The sphere was made from a strongly ferromagnetic material and had a very rough surface. The sphere was magnetically levitated in superfluid $^4$He, and its response with respect to the alternating drive was observed.  At low drives, the velocity response $v$ was proportional to the drive $F_D$, taking the "laminar" form $F_D=\lambda(T) v$, with the temperature-dependent coefficient $\lambda(T)$.  At high drives, the response changed to the "turbulent" form $F_D=\gamma(T) (v^2-v_0^2)$ above the critical velocity $v_0$. At relatively low temperatures the transition from laminar to turbulent response was accompanied by significant hysteresis.

Subsequently, several groups have experimentally investigated the transition to turbulence in superfluids $^4$He and $^3$He-B by using grids \cite{Nichol04a,Nichol04b,Charalambous06,Bradley05b,Bradley06}, wires \cite{Fisher01,Bradley04, Yano07, Hashimoto07, Goto08}, and tuning forks \cite{Blazkova07b}.  The details of the observations are described in a review article \cite{VinenPLTP}. Although a detailed discussion of the cases of $^4$He and $^3$He at both low and relatively high temperatures is required, we shall next briefly describe a few important points.

\begin{figure}[htb] \centering \begin{minipage}[t]{0.95\linewidth} \begin{center} \includegraphics[width=.99\linewidth]{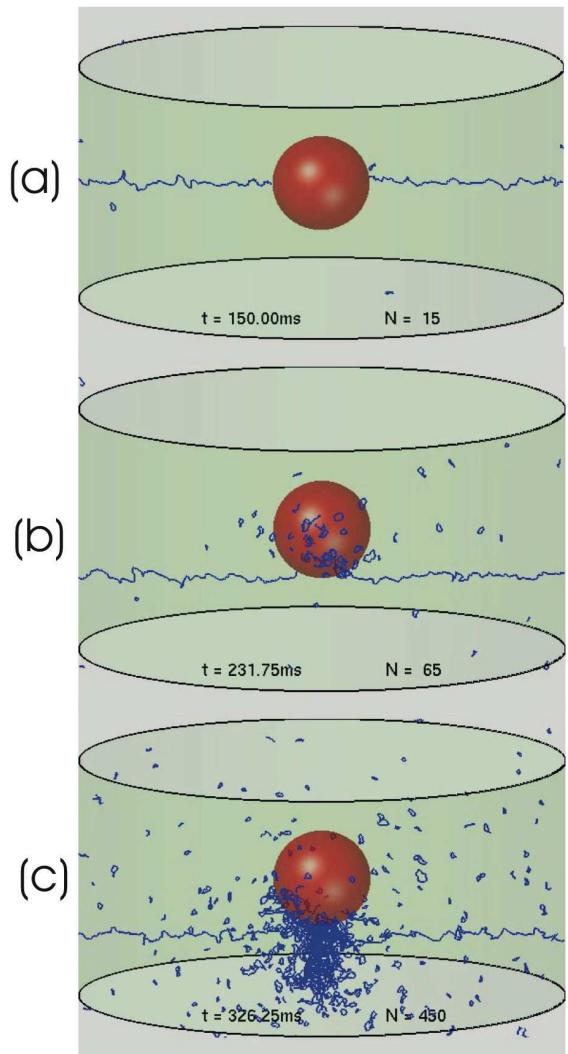} \end{center} \end{minipage} 
\caption{Evolution of the vortex line near a sphere of radius
100~$\mu$m in an oscillating superflow of 150 mm$^{-1}$ at 200~Hz
[H\"anninen, Tsubota and Vinen: Phys. Rev. B \textbf{75} (2007) 064502, reproduced with permission. Copyright 2007 by the American Physical Society].} \label{Hanninen}\end{figure}

These experimental studies reported some common behaviors independent of the details of the structures, such as the type, shape, and surface roughness. The observed critical velocities are in the range from 1 mm/s to approximately 200 mm/s.  Since the velocity is usually much lower than the Landau critical velocity of approximately 50 m/s, the transition to turbulence should come from not intrinsic nucleation of vortices but extension or amplification of remnant vortices. Such behavior is shown in the numerical simulation by the vortex filament model\cite{Hanninen07}.  Figure \ref{Hanninen} shows how the remnant vortices that are initially attached to a sphere develop into turbulence under an oscillating flow. Such behavior must be related to the essence of the observations, many unresolved problems remain, such as the nature of the critical velocity and the origin of the hysteresis for the transition between laminar and turbulent response.

\subsection{Visualization of QT}
There has been little direct experimental information about the flow in superfluid $^4$He. This is mainly because usual flow visualization techniques are not applicable to superfluid. However, the situation is rapidly changing\cite{SciverPLTP}. For QT, we should seed the fluid with tracer particles in order to visualize the flow field and, if possible, quantized vortices, which are observable by appropriate optical techniques.  

A significant contribution was made by Zhang and Van Sciver \cite{Zhang05}. Using the particle image velocimetry (PIV) technique with 1.7-$\mu$m-diameter polymer particles, they visualized a large-scale turbulent flow both in front of and behind a cylinder in a counterflow in superfluid $^4$He at finite temperatures. In classical fluids, such large-scale turbulent structures are seen downstream of objects such as cylinders, with the structures periodically detaching to form a vortex street \cite{Schlichting79}. In the present case of $^4$He counterflow, the locations of the large-scale turbulent structures was relatively stable, and they did not detach and move downstream, although local fluctuations in the turbulence were evident.

Here, it is necessary to understand whether such tracer particles follow the normal flow or the superflow, or even a more complex flow. Poole et al. studied these problems theoretically and numerically to show that the situation changes depending on the size and mass of the tracer particles\cite{Poole05}.

Another significant contribution was the visualization of quantized vortices by Bewley et al. \cite{Bewley06}.  In their experiments, the liquid helium was seeded with solid hydrogen particles smaller than 2.7 $\mu$m at a temperature slightly above $T_\lambda$ after which the fluid was cooled below $T_\lambda$.  When the temperature is above $T_\lambda$, the particles were seen to form a homogeneous cloud that disperses throughout the fluid. However, on passing through $T_\lambda$, the particles coalesced into web-like structures.  Bewley et al. suggested that these structures represent decorated quantized vortex lines. They reported  that the vortex lines appear to form connections rather than remaining separated, and were homogeneously distributed throughout the fluid. The fork-like structures may indicate that several vortices are attached to the same particle, as indicated by numerical simulations of vortex pinning \cite{Tsubota94}. This experimental work on the visualization of quantized vortices is described in detail elsewhere \cite{Paoletti08}. 


\section{Summary and discussions: QT in atomic Bose-Einstein condensates}
In this article, we have reviewed recent research on QT in superfluid helium. Research on QT is currently one of the most important branches in low-temperature physics, attracting the attention of many scientists.  QT is comprised of quantized vortices as definite elements, which differs greatly from conventional turbulence. Thus, investigation of QT may lead to a breakthrough in understanding one of the great mysteries of nature. This last section addresses the topic of QT in atomic Bose-Einstein condensates (BECs).

The realization of Bose--Einstein condensation in trapped atomic gases in 1995 has stimulated intense experimental and theoretical activity in modern physics \cite{Pethick02}.  As proof of the existence of superfluidity, quantized vortices were created and observed in atomic BECs, and numerous efforts have been devoted to solving a number of fascinating problems \cite{KasamatsuPLTP}.  Atomic BECs have several advantages over superfluid helium.  The most important point is that modern optical techniques enable us to directly visualize quantized vortices.   The long history of research into superfluid helium reveals two main cooperative phenomena of quantized vortices. One is a vortex lattice in a rotating superfluid, and the other is a vortex tangle in QT.  To date, most studies of quantized vortices in atomic BECs have been devoted to the former, and few have examined the latter. Recently, it has been shown theoretically that QT can also be created in trapped BECs and that the energy spectrum obeys the Kolmogorov law \cite{Kobayashi07, Kobayashi08}.   Considering the advantages of this system, it would be possible to perform new studies on QT beyond those in superfluid helium.

\section*{Acknowledgment}
The present study was supported in part by a Grant-in-Aid for Scientific Research from JSPS (No. 18340109) and by a Grant-in-Aid for Scientific Research on Priority Areas from MEXT (No. 17071008).

\end{document}